\documentclass[aps,prl,twocolumn,superscriptaddress]{revtex4-1}
\usepackage{amsmath}
\usepackage{graphicx}
\usepackage{epstopdf}
\usepackage[usenames,dvipsnames,svgnames,table]{xcolor}
\usepackage[utf8]{inputenc}
\usepackage{hyperref}
\usepackage{braket}
\usepackage{pbox}
\usepackage{footmisc}
\usepackage{threeparttable}

\newcommand{\AFPl}{\alpha^2F_\text{pl}(\omega)}

\begin{document}

	\title{Plasmonic Superconductivity in Layered Materials}
	
	\author{M. R{\"o}sner}
	\affiliation{Department of Physics and Astronomy, University of Southern California, Los Angeles, CA 90089-0484, USA}
	
	\author{R. E. Groenewald}
	\affiliation{Department of Physics and Astronomy, University of Southern California, Los Angeles, CA 90089-0484, USA}
	
	\author{G. Sch{\"o}nhoff}
	\affiliation{Institut f{\"u}r Theoretische Physik, Universit{\"a}t Bremen, Otto-Hahn-Allee 1, 28359 Bremen, Germany}
	\affiliation{Bremen Center for Computational Materials Science, Universit{\"a}t Bremen, Am Fallturm 1a, 28359 Bremen, Germany}
	
	\author{J. Berges}
	\affiliation{Institut f{\"u}r Theoretische Physik, Universit{\"a}t Bremen, Otto-Hahn-Allee 1, 28359 Bremen, Germany}
	\affiliation{Bremen Center for Computational Materials Science, Universit{\"a}t Bremen, Am Fallturm 1a, 28359 Bremen, Germany}
	
	\author{S. Haas}
	\affiliation{Department of Physics and Astronomy, University of Southern California, Los Angeles, CA 90089-0484, USA}
	
	\author{T. O. Wehling}
	\affiliation{Institut f{\"u}r Theoretische Physik, Universit{\"a}t Bremen, Otto-Hahn-Allee 1, 28359 Bremen, Germany}
	\affiliation{Bremen Center for Computational Materials Science, Universit{\"a}t Bremen, Am Fallturm 1a, 28359 Bremen, Germany}
	
	\date{\today}
	
	\begin{abstract}
		
		Plasmonic excitations behave fundamentally different in layered materials in comparison to bulk systems. They form gapless modes, which in turn couple at low energies to the electrons. Thereby they can strongly influence superconducting instabilities. 

		Here, we show how these excitations can be controlled from the outside via changes in the dielectric environment or in the doping level, which allows for external tuning of the superconducting transition temperature. By solving the gap equation for an effective system, we find that the plasmonic influence can both strongly enhance or reduce the transition temperature, depending on the details of the plasmon-phonon interplay. We formulate simple experimental guidelines to find plasmon-induced elevated transition temperatures in layered materials.

	\end{abstract}

	\maketitle
	
	Due to reduced screening, layered materials promote sizable electron-electron and electron-phonon interactions, leading to a plethora of non-trivial phenomena. In addition to a variety of many-body instabilities, such as charge or superconducting order \cite{berger_strongly_2015,feng_gate-tunable_2015,bradley_characterization_2016,morpurgo_gate-induced_2016}, we also find pronounced many-body excitations such as excitons with binding energies on the eV scale \cite{mak_atomically_2010,bradley_giant_2014} and plasmons with gapless dispersions \cite{liu_plasmon_2008,koppens_graphene_2011,groenewald_valley_2016}. These inherent interactions can be tuned from the outside by adjusting the doping level or the dielectric environment, which in turn affects the aforementioned many-body effects \cite{lin_dielectric_2014,schonhoff_interplay_2016,raja_coulomb_2017,steinhoff_exciton_2017}. The experimental capability to design Van-der-Waals heterostructures in a Lego-like fashion \cite{geim_van_2013} thus promises a fundamentally new opportunity to design material-properties on demand by tuning the interactions from the outside. Critical temperatures for the transition to superconducting or magnetic order might thereby be strongly increased. 
	
	For superconducting properties in layered materials a very promising tuning mechanism of this kind is based on changes to the electron-electron (Coulomb) interaction, since it can be efficiently manipulated from the outside, and it strongly influences the superconducting state. For example, it is well known that the instantaneous Coulomb interaction is repulsive and thus decreases the transition temperature ($T_c$) in conventional superconductors \cite{marques_ab_2005,margine_electron-phonon_2016,akashi_high-temperature_2012}. On the other hand, unconventional paring mechanisms based on the instantaneous interaction can also increase $T_c$ \cite{roldan_interactions_2013}. Furthermore, apart from the instantaneous Coulomb interaction, its dynamic properties also influence the critical temperature. For example, it has been shown that plasmons, which result from the Coulomb interaction dynamics, can affect $T_c$ \cite{takada_plasmon_1978,takada_plasmon_1992,bill_electronic_2003,akashi_development_2013}.
	
	In previous work, we have shown that the tuning mechanism based on changes to the instantaneous Coulomb interaction tends to be rather small \cite{schonhoff_interplay_2016}. Therefore, we focus here on the effects of changes to the dynamic component of the Coulomb interaction and discuss how these can be used to control superconducting properties in layered materials. In particular, we show how plasmonic excitations can be precisely tuned by changing the dielectric environment or doping level of a layered metal, and under which circumstances these excitations strongly couple to the electrons. By solving the gap equation derived from density functional theory for superconductors (SC-DFT), taking into account the conventional electron-phonon interaction along with the static and dynamic Coulomb interaction, we find that these changes to the plasmonic properties indeed significantly influence $T_c$. We demonstrate that the critical temperature can be enhanced as well as reduced due the electron-plasmon coupling. By discussing the subtle interplay between the electron-phonon and the full Coulomb interactions, we finally derive simple design rules for layered materials with pronounced plasmonic enhancements to their critical temperatures.

	\begin{figure}
		\includegraphics[width=0.99\columnwidth]{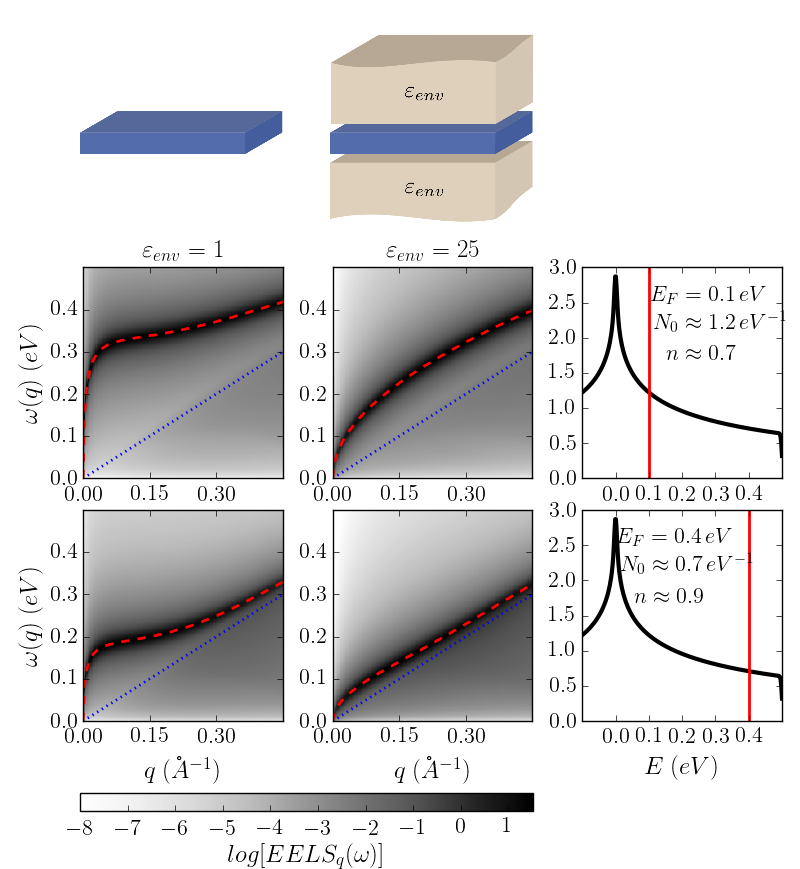}
		\caption{Doping and screening effects to EELS spectra. In the top row we show sketches of a freestanding (left) and dielectrically embedded (right) metallic monolayer. In the second and third row (left columns) we present the EELS spectra as color maps for the different environmental screening situations (left/right panels) and different doping levels (middle/bottom panels). We overlay the EELS spectra with the fitted plasmon frequencies (red dashed lines) and the approximate upper boundary of the particle-hole excitations (blue dotted lines). The shown path starts at $\Gamma$ and continues towards $M$. The right column depicts the density of states per spin for corresponding doping levels of $n \approx 0.7$ with $N_0\approx1.2\,$eV$^{-1}$ and $n \approx 0.9$ with $N_0\approx0.7\,$eV$^{-1}$. \label{fig:EELS}}
	\end{figure}

	\textbf{Results}

	\textbf{Controlling Plasmonic Properties of Layered Metals.}
	The following analysis of plasmonic properties is based on an effective heterostructure defined by a two-dimensional metal with variable doping level placed into a tunable dielectric environment, as depicted in Fig.~\ref{fig:EELS}. All energy scales (band width, Coulomb, and electron-phonon interactions) are chosen comparable to corresponding values found in transition metal dichalcogenides (TMDC) \cite{loon_competition_2017, ritschel_orbital_2015, rossnagel_origin_2011}. We use a two-dimensional (2D) square lattice with doping levels above half filling, nearest-neighbor electron hopping $t = 0.125\,$eV, and lattice constant $a_0=3\,$\AA. The background-screened long-range Coulomb interaction $U_r$ includes effectively the environmental screening effects, rendered by a dielectric constant $\varepsilon_{env}$, and polarization effects from virtual bands, which are otherwise neglected (see Methods for more details). This set of parameters results in a model band width of $D=1\,$eV and a background-screened on-site Coulomb interaction of $U_{r=0} = 1/\Omega \int_{IBZ} dq \, U_q \approx 0.8\,$eV (at $\varepsilon_{env} = 1$), with $\Omega$ being the area of the first Brillouin zone (IBZ) and $U_q$ as defined in the Methods. This is similar to the situation in metallic TMDCs \cite{loon_competition_2017}, although the involved Coulomb interaction is relatively reduced in order to get a moderately correlated model system \cite{ayral_spectral_2012,huang_extended_2014}. The resulting density of states per spin $N_E$ is shown in the right panel of Fig.~\ref{fig:EELS}. In the  half-filled situation ($n=0.5$), the density of states at the Fermi level $N_0$ diverges, while it is monotonously decreasing upon increasing the doping level $n$ towards full filling ($n=1$).

	We examine the plasmonic properties of this model by calculating the electron-energy loss spectra (EELS), 
	\begin{align}
		EELS_q(\omega) = -\operatorname{Im}
						   \left(
							\frac{1}{\varepsilon_q(\omega)}
						   \right) \label{eq:EELS}
	\end{align}
	using the random phase approximation (RPA) to evaluate the polarization function $\Pi_q(\omega)$ as needed in the dielectric function,
	\begin{align}
		\varepsilon_q(\omega) = 
			1 + U_q \Pi_q(\omega).
	\end{align}
	Here, $U_q$ is the Fourier transform of $U_r$ (see Methods for details). In Fig.~\ref{fig:EELS} we show the resulting spectra for two doping levels ($n \approx 0.7$ and $n\approx0.9$ corresponding to $N_0 \approx 1.2\,$eV$^{-1}$ and $N_0\approx0.7\,$eV$^{-1}$, respectively) and two environmental screenings ($\varepsilon_{env}=1$ and $\varepsilon_{env}=25$). We indicate the upper boundary of the particle-hole continuum by blue dotted lines. Next to this continuum, we can clearly identify a well defined strong resonance (highlighted in red), which follows the plasmonic dispersion $\omega_q^{pl}$ defined by $\operatorname{Re}\left[\varepsilon(q, \omega_q^{pl})\right] = 0$. In all of these cases we find gapless modes with $\sqrt{q}$-like dispersions for small momenta, which flatten at intermediate $q$ before smoothly hybridizing with the particle-hole continuum, which is characteristic for plasmons in 2D metals \cite{groenewald_valley_2016,liu_plasmon_2008,hwang_dielectric_2007}. Most importantly, Fig.~\ref{fig:EELS} illustrates the sensitivity of the plasmon dispersion to the layer's environment and to the doping level. Increasing the environmental screening from $\varepsilon_{env} = 1$ to $25$ decreases the dispersion for $q \lesssim 0.25\,$\AA$^{-1}$ only, which directly results from the decreased Coulomb interaction $U_q$ in this range (see Methods). A similar effect has been predicted for heterostructures consisting of graphene and hexagonal boron nitride\cite{andersen_dielectric_2015}. Increasing the electron-doping level from $n\approx0.7$ to $n\approx0.9$ induces a reduction of the plasmon frequencies for the entire $q$-range. This results from the density of states at the Fermi level being reduced from $N_0\approx1.2\,$eV$^{-1}$ to $N_0\approx0.7\,$eV$^{-1}$, which directly influences the plasmonic dispersion as we describe in more detail in the Methods section. 
	
	To understand how tuning $n$ and $\varepsilon_{env}$ influences the electronic and superconducting properties, we analyze the anomalous electronic self energy in the Nambu space \cite {nambu_quasi-particles_1960} and within the $GW$ approximation using a plasmon pole fit to the full dynamic Coulomb interaction, yielding
	\begin{align}
		\Sigma_k^{dyn}(i\omega_n) = 
			\frac{1}{\beta}
			\sum_{k'm} 
				G_{k}(i\omega_m) |a_q^{pl}|^2 D_q^{pl}(i\omega_n-i\omega_m)
	\end{align}
	as the only dynamic contribution (see Methods). Here, $G_k$ and $D_q^{pl}$ are the electronic and plasmonic (bosonic) propagators, and $a_q^{pl}$ is the electron-plasmon coupling with $q = k-k'$. This electron-plasmon self energy is fully equivalent to the anomalous electron-phonon self energy involved in conventional superconductivity theory\footnote{Note that the self consistency in Eliashberg theory updates the electronic propagator, while in SC-DFT the electronic propagator is fixed to its initial value $G_0$  \cite{sanna_introduction_2017}.}. Therefore, we can generalize the standard phonon-based formalism \cite{allen_theory_1983} and introduce a Fermi surface averaged \emph{plasmonic Eliashberg function}
	\begin{align}
		\alpha^2F_{pl}(\omega) =	
			\frac{1}{N_0}
			\sum_{kk'}
			\delta(\xi_k) 
			\delta(\xi_{k'})
			\delta(\omega^{pl}_{k-k'} - \omega)
			a_{k-k'}^{pl},
	\end{align}
	which describes the average efficiency of plasmons with frequency $\omega$ to scatter electrons from any state near the Fermi surface to any other near the Fermi surface. Based on $\alpha^2F_{pl}(\omega)$, we can furthermore define the effective electron-plasmon coupling constant,
	\begin{align}
		\lambda^{pl} &= 2\int \frac{\alpha^2F^{pl}(\omega)}{\omega} d\omega,
		\label{eq:LambdaPlasmon}
	\end{align}
	as well as the effective plasmonic frequency,
	\begin{align}
		\omega^{pl}_{log} &= 
		\operatorname{exp}
		\left[
		\frac{2 \int \frac{\alpha^2F^{pl}(\omega)}{\omega} \log{\omega} \,d\omega}{\lambda^{pl}}
		\right].
	\end{align}
	\begin{figure}
		\includegraphics[width=0.999\columnwidth]{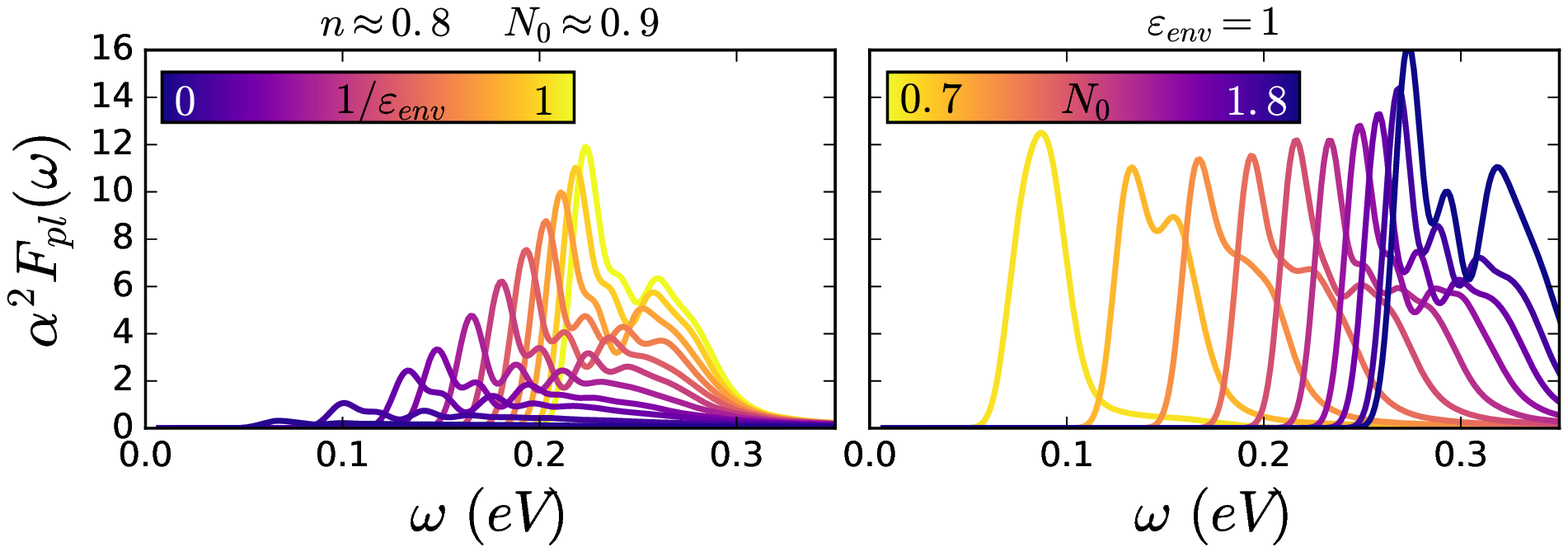}\\
		\vspace{-0.35cm}
		\includegraphics[width=0.999\columnwidth]{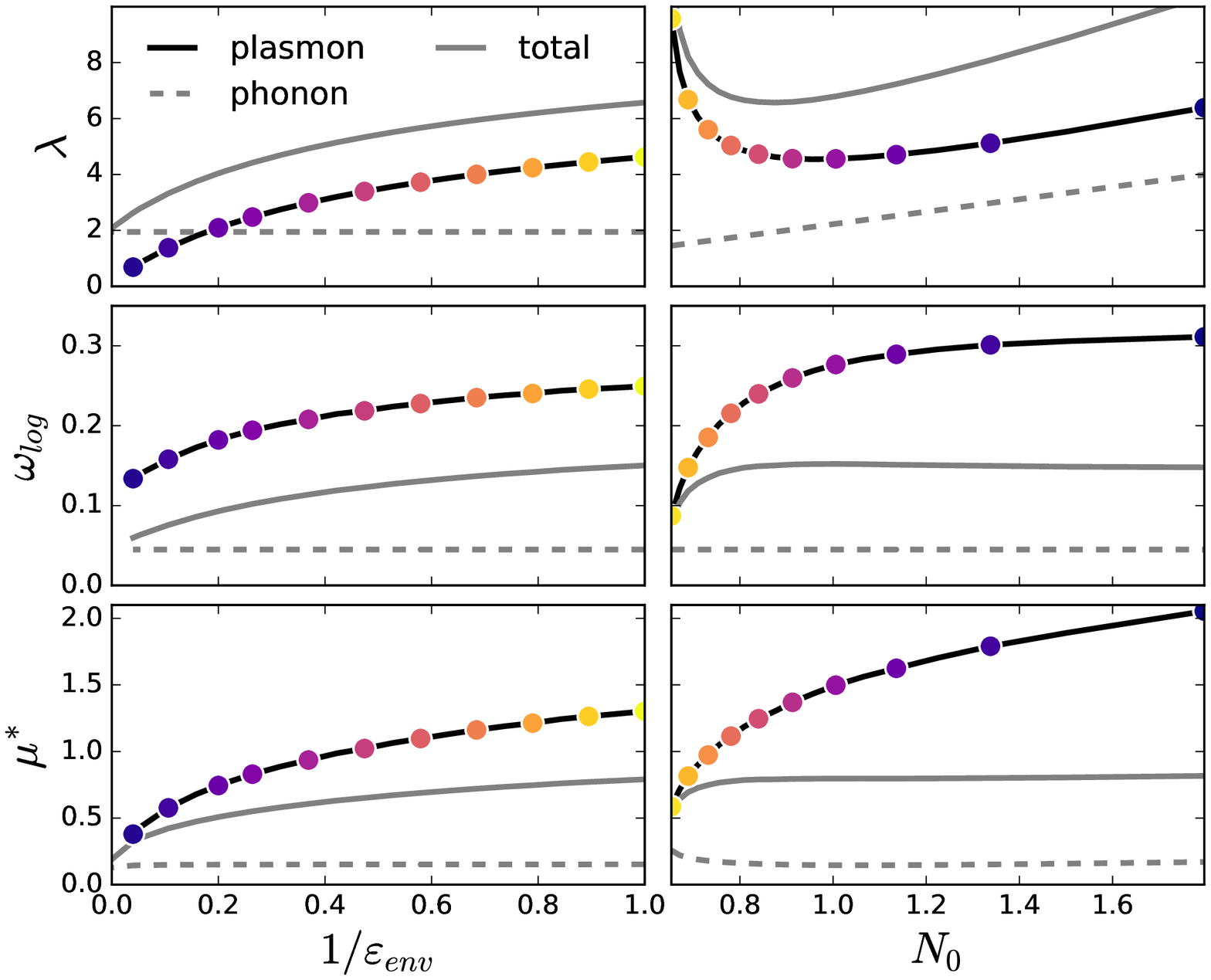}
		\caption{Plasmonic properties. Colorized data represents Plasmonic Eliashberg functions $\alpha^2F_{pl}(\omega)$, effective electron-plasmon couplings $\lambda^{pl}$, effective plasmonic frequencies $\omega^{pl}_{log}$, and effective static repulsions $\mu^{pl*}$ for different dielectric environments (left panels) and doping levels (right panels). For the left panels we used $n \approx 0.8$ ($N_0 \approx 0.9\,$eV$^{-1}$) and $\varepsilon_{env}=1$ for the right panels. The colors of the $\alpha^2F_{pl}(\omega)$ plots correspond to the different values of $N_0$ and $\varepsilon_{env}$. Dashed (solid) gray lines correspond to the phononic (total) parameters using a constant frequency $\omega^{ph}=45\,$meV and an electron-phonon coupling matrix element $g^2=50\,$meV$^2$.  \label{fig:ElPlCoup}}
	\end{figure}%
	In Fig.~\ref{fig:ElPlCoup} all of these quantities \footnote{
			Note that these parameters quantify the pure plasmonic effects which is different to the approach by Das and Dolui \cite{das_superconducting_2015} where the plasmonic renormalization of the \emph{phononic} Eliashberg function is evaluated. 
	} are shown as functions of the environmental screening $\varepsilon_{env}$ and of the doping level $n$. From here on, we describe the doping level by the resulting density of states at the Fermi level $N_0$ (and not $n$) due to its direct influence to plasmonic properties.

	In the case of varying $\varepsilon_{env}$ (left panels) we find the peaks in $\alpha^2F_{pl}(\omega)$ shifting to lower energies and decreasing in amplitude with increasing environmental screening (decreasing $1/\varepsilon_{env}$). This shift in energy results from the general decrease of the plasmon frequencies with increasing screening as already observed in Fig.~\ref{fig:EELS}. The strongly decreasing peak height originates from the electron-plasmon coupling which is approximately given by $|a_q^{pl}|^2 \propto \omega_q^{pl} U_q$. The coupling is thus reduced due the decreasing frequencies $\omega_q^{pl}$ and, most importantly, due to $U_q$, which is strongly reduced by $\varepsilon_{env}$ (see Methods). These trends are also observed in the effective electron-plasmon coupling $\lambda^{pl}$ and in the effective plasmon energy $\omega_{log}^{pl}$, which both decrease with increasing $\varepsilon_{env}$. 
	
	For the case of varying doping level $n$ and thus changing $N_0$ (right panels) we primarily find a shift of $\alpha^2F_{pl}(\omega)$ to higher energies when increasing $N_0$, without any major changes in its spectral weight. The shift is again explained by the raised plasmon frequencies. At the same time the electron-plasmon coupling $|a_q^{pl}|^2$ is increased due to its $\omega_q^{pl}$ dependence. Thus, $|a_q^{pl}|^2$ actually increases the peak height of $\alpha^2F_{pl}(\omega)$ for higher $N_0$. However, the pure plasmon density of state $F_{pl}(\omega) = \sum_q \delta(\omega_{q}^{pl} - \omega)$ shifts to higher frequencies and looses weight for small frequencies (see Supplement) which compensates for the enhancement from $|a_q^{pl}|^2$ for larger $N_0$ to $\alpha^2F_{pl}(\omega)$ yielding just slightly elevated $\lambda^{pl}$ for comparably large $N_0$. The largest effective plasmonic couplings are found for small $N_0$. This results from the constant shift of $\alpha^2F_{pl}(\omega)$ to small frequencies with decreasing $N_0$ (which in turn results from a strongly enhanced $F_{pl}(\omega)$ for small $N_0$ and small $\omega$) together with the definition in Eq.~(\ref{eq:LambdaPlasmon}) which favors small frequencies. For intermediate $N_0$, the effective plasmon frequency is comparably large, while $|a_q^{pl}|^2$ is not big enough to compensate for the reduced plasmonic density of states, yielding a reduced effective coupling $\lambda^{pl}$ here. 
	
	\textbf{Plasmonic Superconductivity.} After discussing the plasmonic characteristics, we now turn to their influence on superconducting properties. Since the electron-plasmon coupling matrix element $a^{pl}_q$ is rather isotropic, we neglect the superconducting gap anisotropy over the Fermi surface and assume $s$-wave symmetry of the latter \cite{allen_theory_1983}. Under this assumption, we can solve the linearized SC-DFT gap equation in energy space \cite{luders_ab_2005, marques_ab_2005},
	\begin{align}
		\Delta(\xi) =
		& - Z(\xi) \Delta(\xi) \label{eq:Gap} \\
		& - \frac{1}{2} \int d\xi' 
		N(\xi')
		K(\xi,\xi')	
		\frac{\operatorname{tanh}\left[ (\beta/2) \xi' \right]}{\xi'}
		\Delta(\xi'), \notag
	\end{align}
	where $\xi$ and $\xi'$ are energies, $\Delta(\xi)$ is the SC-DFT gap function, $Z(\xi)$ is the renormalization factor, $K(\xi,\xi')$ is the kernel, and $\beta$ is the inverse temperature. The critical temperature $T_c$ is defined as the maximum temperature for which a non-trivial solution to this equation exists.
	
	For a conventional superconductor, we need to take into account the electron-phonon coupling induced mass renormalization factor $Z^{ph}$ and the corresponding kernel $K^{ph}$ as well as the contribution of the static Coulomb interaction $K^{stat}$ ($K = K^{ph} + K^{stat}$). While the latter can be readily approximated as a constant $N_0 K^{stat} = \mu$ \cite{akashi_high-temperature_2012}, $Z^{ph}$ and $K^{ph}$ are non-trivial functions of $\xi$ and $\xi'$, as discussed in the Methods section. $Z^{ph}$ and $K^{ph}$ have, however, simple limits for zero temperature: $Z^{ph}(0) = \lambda^{ph}$ and $N_0 K^{ph}(0,0) = -\lambda^{ph}$ \cite{luders_ab_2005}. 
	 
	If we use these limits to approximate $Z^{ph}(\xi)$ and $K^{ph}(\xi,\xi')$ within a phononic energy window $\omega^{ph}$, in which these functions are non-zero, we can solve the gap equation analytically and get an explicit expression for the critical temperature \cite{akashi_density_2014,luders_ab_2005,morel_calculation_1962},
	\begin{align}
		T_c \propto \omega^{ph} 
		\operatorname{exp}
		\left[ 
		-\frac{1 + Z^{ph}}{K^{ph} - \mu^{ph*}}
		\right]
		\propto
		\omega^{ph}
		\operatorname{exp}
		\left[ 
		-\frac{1 + \lambda^{ph}}{\lambda^{ph} - \mu^{ph*}}
		\right]
		,
		\label{eq:TcSimple}
	\end{align}
	where $\mu^{ph*} = \mu / (1 + \operatorname{ln}(E_F/\omega^{ph}))$ is the retarded Morel-Anderson pseudo potential. From this expression and Eq.~(\ref{eq:Gap}) we get a qualitative understanding of all parameters: $\lambda^{ph}$ enhances the gap and thus $T_c$ as a result of the attractive phononic kernel, but also reduces it due to the mass-renormalization. The static Coulomb interaction, as rendered by $\mu^{ph*}$, always lowers $T_c$. 
	
	In order to additionally include dynamic Coulomb effects and thus plasmonic contributions, we make use of the dynamic kernel $K^\text{dyn}_{k,k-q}$ defined by Akashi and Arita \cite{akashi_development_2013} and combine it with the single plasmon pole approximation (see Methods). Afterwards we evaluate its Fermi surface average and obtain
	\begin{align}
		& K^{dyn}(\xi,\xi') =
			\frac{1}{N_\xi N_{\xi'}}
		    \sum_{k,q}
		      \delta(\xi-\xi_k) 
		      \delta(\xi'-\xi_{k-q}) 
		      K^\text{dyn}_{k,k-q} \\
		&\approx \frac{2}{N_0} \int d\omega \, \AFPl \,
			  \left(
			  \frac{1}{\omega} +
			  \frac{
				I_K(\xi, \xi', \omega)
				-
				I_K(\xi,-\xi', \omega)
			  }{
				\operatorname{tanh}\left[ (\beta/2) \xi \right]
				\operatorname{tanh}\left[ (\beta/2) \xi' \right]
			  }
			  \right). \notag
	\end{align}
	If we additionally make use of the $\lambda^{pl}$ definition given in Eq.~(\ref{eq:LambdaPlasmon}), we can rewrite the dynamic kernel,
	\begin{align}
			N_0 K^{dyn}(\xi, \xi') = 	
			\lambda^{pl}+ N_0 \Delta K^\text{dyn}(\xi, \xi'), \label{eq:KDyn}
	\end{align}
	where we absorbed all dynamics into the second term, which behaves exactly like the phononic $K^{ph}(\xi, \xi')$, i.e., in the static limit it becomes $N_0 \Delta K^\text{dyn}(0, 0) = -\lambda^{pl}$ and vanishes for large $\xi$ and $\xi'$. The complete dynamic Coulomb Kernel thus vanishes for $\xi = \xi' = 0$ and is strictly positive otherwise (see Supplement). The full kernel then reads $K = K^{ph} + K^{stat} + K^{dyn}$. 
	
	We numerically solve the SC-DFT gap equation, including fully energy dependent expressions for $Z(\xi)$ and $K(\xi,\xi')$ subject to the influence of different doping levels and varying dielectric environments. For the phononic contribution, we use a simple Einstein mode of frequency $\omega^{ph}$ and constant coupling $g^2$. The Coulomb contributions are directly derived from the model. 

	\begin{figure}
		\includegraphics[width=0.999\columnwidth]{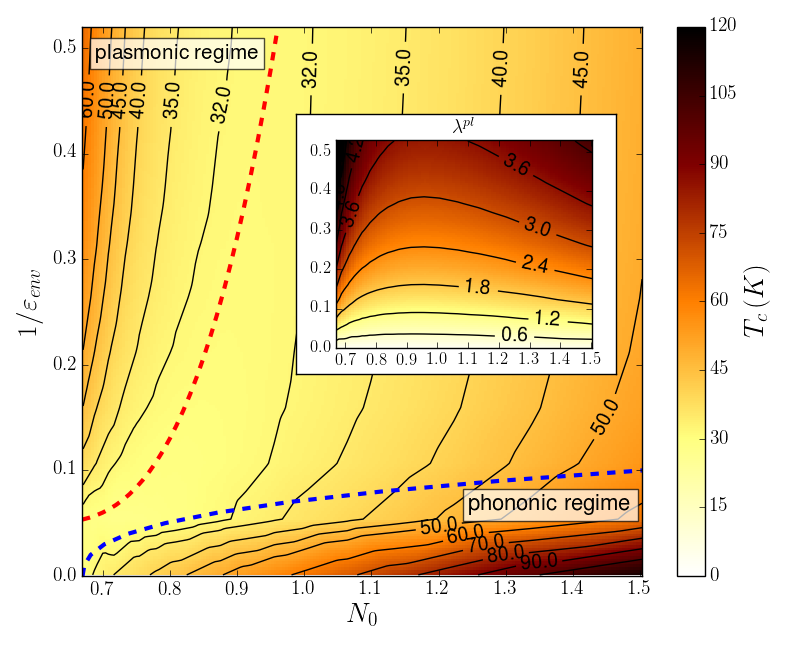}
		\caption{$T_c$ as a function of doping (rendered by $N_0$) and environmental screening, including the effects of phonons, plasmons, and static Coulomb interaction. Solid black lines correspond to constant values of $T_c$. ``Phononic'' and ``plasmonic'' regimes are marked by blue and red dotted lines, respectively. The inset depicts the effective electron-plasmon coupling $\lambda^{pl}$. \label{fig:MapNEps}}
	\end{figure}

	In Fig.~\ref{fig:MapNEps} we show resulting $T_c$ for electron-doping levels increasing from $n \gtrsim 0.5 $ to $1.0$ (corresponding to $N_0\approx1.5\,$eV$^{-1}$ to $0.7\,$eV$^{-1}$) and for varying dielectric environments. The phononic properties are set to $\omega^{ph}= 40\,$meV (similar to the optical modes in TMDCs \cite{heil_origin_2017}) and $g^2 = 45\,$meV$^2$ yielding realistic a $\lambda^{ph}\approx1.5-3.5$. Both, the doping level and the dielectric environment, affect the critical temperature. We find enhanced $T_c$ for large $N_0$ and $\varepsilon_{env}$ as well as for small $N_0$ and $\varepsilon_{env}$. These regimes are labeled as ``phononic'' and ``plasmonic'', respectively. In the phononic regime, $\lambda^{pl}$ is relatively small (see inset of Fig.~\ref{fig:MapNEps}), while $\lambda^{ph} = 2 N_0 g^2 / \omega^{ph}$ constantly increases with $N_0$. The increasing trend of the critical temperature thus follows the effective phononic coupling in this regime. In the plasmonic regime, $\lambda^{pl}$ is large and increases with decreasing environmental screening. Since $\varepsilon_{env}$ has no effect on the phononic properties in our model, the increasing $T_c$ follows $\lambda^{pl}$. For large $N_0$ and small $\varepsilon_{env}$, and thus rather large electron-plasmon and electron-phonon couplings, we also find enhanced critical temperatures. Although the total coupling seems to be the strongest here, we do not find the highest critical temperatures in this regime. 
	
	\begin{figure}
		\includegraphics[width=0.99\columnwidth]{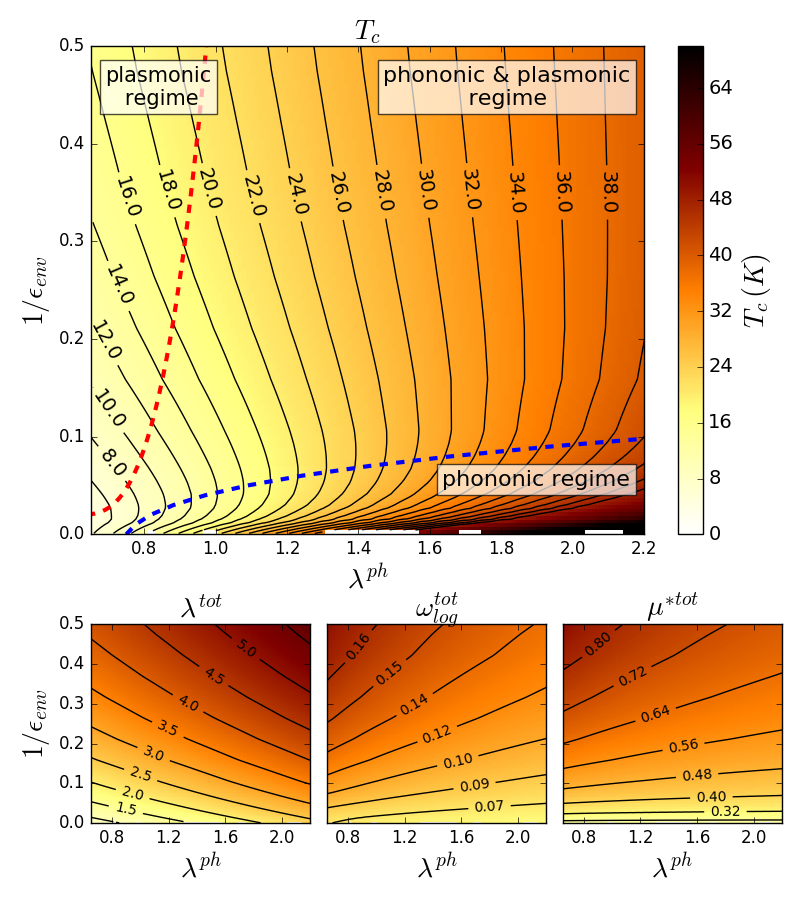}
		\caption{$T_c$ as a function of phononic coupling and environmental screening. The upper panel shows the full critical temperature including the effects of phonons, plasmons and static Coulomb interaction. The lower panels show the total effective coupling $\lambda^{tot}$, total effective frequency $\omega^{tot}_{log}$, and total effective Coulomb repulsion $\mu^{*tot}$. \label{fig:MapEpsPh}}
	\end{figure}
	
	In order to understand this interplay between the two coupling channels in more detail, we now fix the doping level to $n \approx 0.8$ ($N_0 \approx 0.9\,$eV$^{-1}$) and study the effects of simultaneously varying electron-phonon coupling $g^2$ and environmental screening $\varepsilon_{env}$. The former now solely controls $\lambda^{ph}$ and the latter $\lambda^{pl}$. In Fig.~\ref{fig:MapEpsPh} we show results for $\omega^{ph} = 60\,$ meV and $g^2 \approx 22 - 76\,$meV$^2$, yielding $\lambda^{ph} \approx 0.65 - 2.2$. Next to $T_c$, we also show the total effective coupling $\lambda^{tot}$, frequency $\omega^{tot}$ and static repulsion $\mu^{*tot}$ defined by
	\begin{align}
		\lambda^{tot} &= \lambda^{ph} + \lambda^{pl} \\
		\omega^{tot} &= \operatorname{exp} 
						\left[ 
						\frac{
							\operatorname{log}(\omega^{ph}) \lambda^{ph}
							+ 
							\operatorname{log}(\omega^{pl}) \lambda^{pl}
						}{\lambda^{ph} + \lambda^{pl}}
						\right] \\
		\mu^{*tot} &= \frac{\mu^{tot}}{1 + \operatorname{ln}(E_F/\omega^{tot})}
		\quad\text{ with }\quad
		\mu^{tot} = \mu + \lambda^{pl}.
	\end{align}
	It is important to note that $\lambda^{pl}$ needs to be taken into account in both, $\lambda^{tot}$ and $\mu^{tot}$. This is a direct result from the two terms of the dynamic Coulomb contribution defined by Eq.~(\ref{eq:KDyn}). The first (constant) term contributes to $\mu^{tot}$, which reduces $T_c$, and the second (dynamic) term to $\lambda^{tot}$, which enhances $T_c$. In Fig.~\ref{fig:ElPlCoup} we show how these total effective parameters depend on the environmental screening and the doping level. Most importantly, we see here that the total effective coupling is simply enhanced, while the total effective frequency and static repulsion are strongly reduced compared to the pure plasmonic quantities.
	
	Based on the interplay of these three parameters, we can now qualitatively understand the behavior of $T_c$ shown in Fig.~\ref{fig:MapEpsPh}. In the phononic regime neither the total effective frequency nor the static repulsion change drastically with increasing $\lambda^{ph}$ (see lower panels in Fig.~\ref{fig:MapEpsPh}). Only the total effective coupling increases with $\lambda^{ph}$, which is responsible for the increasing $T_c$ trend, here. Similarly, in the plasmonic regime $T_c$ increases towards small environmental screenings due to the enhancement of $\lambda^{tot}$ and $\omega^{tot}_{log}$, which is driven by the increasing trend of the plasmonic $\lambda^{pl}$ and $\omega^{pl}_{log}$, respectively. Here, however, $\mu^{tot}$ and $\mu^{*tot}$ also increase with $\lambda^{pl}$, which reduces the increasing trend in $T_c$. In the remaining regime the total effective coupling $\lambda^{tot}$ is enhanced by both coupling channels. Interestingly, $T_c$ seems to be mostly controlled by $\lambda^{ph}$ here (lines of constant $T_c$ are vertical), whereas $\varepsilon_{env}$ and thus $\lambda^{pl}$ seem to have a negligible effect. However, by studying the effective total parameters, we realize that $T_c$ is simultaneously enhanced and reduced by the counteracting trends in $\lambda^{tot}$ and $\mu^{*tot}$ with increasing $\lambda^{pl}$. It is thus an interplay between both coupling channels which is responsible for $T_c$ here.

	\begin{figure}
		\includegraphics[width=0.99\columnwidth]{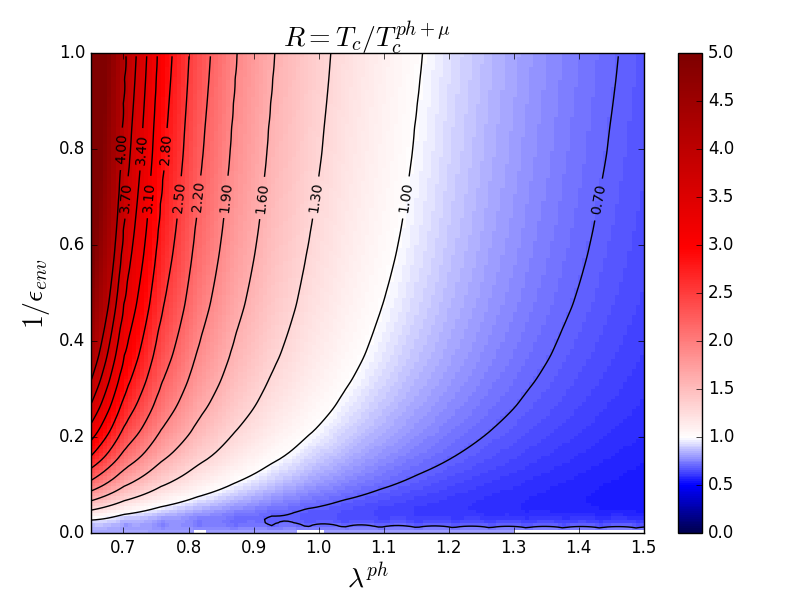}
		\caption{Plasmonic enhancement-reduction map. We show the ratio of the full transition temperature $T_c$ and the phononic one $T_c^{ph+\mu}$ (excluding plasmonic effects). Red parts represent the regime of plasmonic enhancement and blue the pasmonic reduction. \label{fig:MapR}}
	\end{figure}	

	Finally, we draw our attention to the ratio map in Fig.~\ref{fig:MapR}, which depicts $R = T_c / T_c^{ph+\mu}$, where $T_c^{ph+\mu}$ is the critical temperature including only phononic and static Coulomb effects (no plasmonic contributions). This quantity reveals those regimes where the plasmons increase $T_c$ ($R > 1$) in comparison to the situation without any plasmonic influence ($R=0$) and those regimes where the plasmons reduce it ($R<1$). We find a narrow regime of plasmonic enhancement for most dielectric screenings and rather small phononic couplings. The amount of the plasmonic enhancement within this stripe is controlled by the environmental screening. In the freestanding situation ($\varepsilon_{env}=1$), we find full critical temperatures which are enhanced by factors up to $5$. By increasing the environmental screening, we decrease this enhancement. 
	
	If we increase the phononic coupling we also find a decreasing plasmonic enhancement, which is in line with previous ab initio data by Akashi and Arita \cite{akashi_development_2013}. If we further increase the phononic coupling we arrive in a situation with plasmonic reduction, which holds for all environmental screenings. Similar charge-fluctuation induced reductions of the critical temperature has also been reported in Ref.~\cite{das_superconducting_2015}. We can get an universal understanding of this quite complex behavior by using the approximate model function for $T_c$ from Eq.~(\ref{eq:TcSimple})
	\begin{align}
		R = \frac{T_c}{T_c^{ph+\mu}}
		  = \frac{\omega^{tot}}{\omega^{ph}}
		    \operatorname{exp}
				\left[ 
					\frac{1+\lambda^{ph}}{\lambda^{ph} - \mu^{ph*}}
				   -\frac{1+\lambda^{tot}}{\lambda^{tot} - \mu^{tot*}}
				\right].
	\end{align}
	From this, we can identify two factors controlling $R$. The first one is the ratio of the involved frequencies. Since this is always bigger than $1$ (see $\omega^{tot}$ map in Fig.~\ref{fig:MapEpsPh} and remember $\omega^{ph} = 60\,$meV) we can rule it out for being responsible for the reduction. Thus, the exponential term must be responsible for the plasmonic reduction. And indeed, as we show in the Supplemental Material, in the limit $\lambda^{ph} \gg \lambda^{pl}$ it is easy to show, that this term becomes smaller than $1$ if $\mu^{tot*} >\mu^{ph*}$ holds, which is true for large $\lambda^{ph}$ and $\omega^{tot}>\omega^{ph}$. 
	
	The plasmonic reduction for enhanced phononic coupling is therefore a result of the plasmonic enhancement of the pseudo Coulomb potential $\mu^{tot*}$, whereas the plasmonic contribution to the total effective coupling becomes negligible and $\lambda^{tot} \approx \lambda^{ph}$. On the other side, if the phononic coupling is weak, the plasmonic contribution to the total effective coupling becomes relevant. Here, the effective coupling can overcome the effective Coulomb potential leading to a net enhancement of $T_c$ in comparison to $T_c^{ph+\mu}$. It is thus a very subtle interplay between all interactions which controls the resulting transition temperature.
	
	\textbf{Discussion} 

	We have shown how the plasmon frequencies and the electron-plasmon coupling of a layered metal can be tuned by electron doping and environmental screening. While the latter always reduces both, the plasmon frequencies and the couplings, the former has more subtle consequences. Most importantly, we found that the coupling is enhanced while the frequency is reduced for lowered density of states at the Fermi level. These strongly screening and doping dependent plasmonic properties imprint strong changes to the full superconducting transition temperature, allowing for external control of the latter. Thereby, we demonstrated that the electron-plasmon coupling does not only contribute to the total attractive coupling, but also to the total repulsive static Coulomb term. As a consequence, there is both plasmonic enhancement and reduction depending on the doping level and dielectric environment.

	In order to find a sweet spot for plasmon-enhanced superconductivity in an experiment, the major task is to enhance the effective electron-plasmon coupling $\lambda^{pl}$ while trying to keep the effective plasmon frequency $\omega^{pl}$ low. The latter is important since $\lambda^{pl}$ necessarily also adds to the effective Coulomb repulsion $\mu^{ph*}$, which is, however, decreased by a decreasing $\omega^{pl}$. These properties are indeed not contradicting as evident from Eq.~(\ref{eq:LambdaPlasmon}): an enhanced plasmonic spectral weight around small frequencies enhances the effective coupling. Based on these considerations we can define two rules:
	\begin{itemize}
		\item[(I)] Use two-dimensional metals with gapless plasmonic modes.
		\item[(II)] Use metals with a \emph{reduced} density of states at the Fermi level, i.e. with a small effective mass.
	\end{itemize}
	Both rules guarantee a decreased $\omega^{pl}$ and an enhanced $\lambda^{pl}$. In order to further enhance the effective plasmonic coupling we can define two additional rules:
	\begin{itemize}
		\item[(III)] Reduce the environmental screening.
		\item[(IV)] Avoid inter-band polarization effects to reduce Landau damping, i.e. try to find a single free-standing metallic band embedded in an electronic band gap.
	\end{itemize}
	The former actually increases $\omega^{pl}$ which, however, can be compensated by a strongly increased $\lambda^{pl}$ at small $\lambda^{ph}$ (see top left corner of the ratio map in Fig.~\ref{fig:MapR}). The latter is discussed in Ref. \onlinecite{gjerding_band_2017}. This also leads to the last rule:
	\begin{itemize}
		\item[(V)] Use a material with small intrinsic electron-phonon coupling.
	\end{itemize}
	
	Promising candidates which fulfill at least some of these guidelines include slightly electron or hole doped semiconducting TMDC monolayers \cite{ge_phonon-mediated_2013, rosner_phase_2014} (metallic TMDCs might show too strong electron-phonon interactions \cite{suderow_pressure_2005}, too strong inter-band polarizations \cite{andersen_plasmons_2013}, and too high density of states at the Fermi level), monolayers of recently proposed 1T-AlCl$_2$ \cite{gjerding_band_2017}, and in general singlelayer $sp$-electron systems, such as hexagonal boron nitride or functionalized graphene \cite{miro_atlas_2014, karlicky_halogenated_2013}.

	In order to disentangle plasmonic and phononic effects from each other, it would be best if the effects of both, the environmental screening and the doping level, could be experimentally studied. If there are strong changes in $T_c$ by changing the dielectric environment or if $T_c$ increases by decreasing the density of states at the Fermi level, our results show that it is very likely that $T_c$ is significantly controlled by the coupling between electrons and plasmons.
	
	These findings clearly show that static and dynamic Coulomb interaction effects need to be accurately considered in order to explain superconducting properties from a theoretical point of view. At the same time they point towards exciting new directions in the field of on-demand material-property design using layered systems. Here, a sophisticated choice of materials can increase critical temperatures by combining advantageous properties from different materials.

	\textbf{Methods}
	
	\textbf{Realistic Coulomb Interactions.} For a realistic description of the Coulomb interactions, we imagine the metallic band to be part of a multi-band structure, e.g. formed by $s$, $p$, and $d$ orbitals. While we concentrate in the main text on the \emph{low-energy} subspace around the Fermi level, it is important to realize that the neglected high-energy parts of the band structure have a screening influence on the Coulomb interaction in the low-energy subspace \cite{aryasetiawan_frequency-dependent_2004}. If the metallic band from the main text is the only band crossing the Fermi level, we can readily approximate the polarization function of the neglected or remaining part of the band structure as a static function $\Pi^{rest}_q$. The total polarization function then reads $\Pi^{total}_q(\omega) = \Pi_q(\omega) + \Pi^{rest}_q$, and the full dynamic Coulomb interaction is given by
	\begin{align}
		W_q(\omega) &= \frac{v_q}{1 - v_q \left[ \Pi_q(\omega) + \Pi^{rest}_q \right]} \notag \\
					 &= \frac{U_q}{1 - U_q \Pi_q(\omega)}
					  = \frac{U_q}{\varepsilon_q(\omega)}, \label{eq:WFull}
	\end{align}
	where $v_q$ is the bare interaction, and $U_q$ the background-screened interaction is defined by
	\begin{align}
		U_q = \frac{v_q}{1 - v_q \Pi^{rest}_q}
		     = \frac{v_q}{\varepsilon^{rest}_q}.
	\end{align}
	The background dielectric function $\varepsilon_{rest}(q)$ now renders all screening effects resulting from the neglected part of the band structure and also those resulting from the dielectric environment. This function can be derived from classical electrostatics. In the case of a layered system with thickness $d$ and embedded between two semi-infinite dielectric substrates with dielectric constants $\varepsilon_{env}$ it reads \cite{keldysh_coulomb_1979, rosner_wannier_2015}
	\begin{align}
		\varepsilon_{rest}(q) = 
			\varepsilon_\infty
			\frac{1 - \tilde{\varepsilon}^2 e^{-2qd}}
			     {1 + 2\tilde{\varepsilon} e^{-qd} + \tilde{\varepsilon}^2 e^{-2qd}}, \label{eq:EpsRest}
	\end{align}
	with
	\begin{align}
		\tilde{\varepsilon} = \frac{\varepsilon_\infty - \varepsilon_{env}}{\varepsilon_\infty + \varepsilon_{env}}.
	\end{align}
	This function smoothly interpolates between the long-wavelength limit $\varepsilon_{rest}(q \rightarrow 0) = \varepsilon_{env}$ and the short-wavelength limit $\varepsilon_{rest}(q \rightarrow \infty) = \varepsilon_\infty$ of the layered system.	Thus, $U_q$ is mostly affected by $\varepsilon_{env}$ for small momenta, while the internal $\varepsilon_\infty$ controls short wavelengths.
	
	If we additionally define the bare interaction as $v_q = 2\pi e^2 / [A(q + \gamma q^2)]$, where $A$ is the unit cell size and $\gamma$ is an effective form-factor rendering effects from the non-zero height of the layer, the realistic background-screened Coulomb interaction is fully defined by the parameters $A$, $\gamma$, $\varepsilon_\infty$, and $d$.
	
	To obtain the fully screened Coulomb interaction $W_q(\omega)$, we need to evaluate Eq.~(\ref{eq:WFull}). Therefore, we utilize the polarization function of the metallic band  in its random phase approximation, which is given by  
	\begin{align}
		\Pi_q(\omega) = 
			\frac{2}{N_k}
			\sum_k
				\frac{f(k-q) - f(k)}{\xi(k-q) - \xi(k) + \omega + i0^+},
	\end{align}
	where $N_k$ is the number of involved $k$ points (of the entire Brillouin zone), $2$ is the standard spin factor, $f(k)$ is the Fermi distribution function, $\xi(k)$ is the spin-degenerated electronic dispersion of the system, and $i0^+$ is an infinitesimal positive imaginary number \cite{groenewald_valley_2016}. 
	
	\textbf{Plasmon Pole Model and Fitting.} We define the single plasmon-pole model via 	
	\begin{align}
	W_{q}(\omega) \approx \
		W_{q}(0)
		+ 2 |a_q^{pl}|^2
		\left( 
		\frac{1}{\omega_q^{pl}}
		+ \frac{\omega_q^{pl}}{\omega^2 - (\omega_q^{pl})^{2}}
		\right), \label{eq:PPM}
	\end{align}
	where $W_{q}(0)$ is the static screened Coulomb interaction, $\omega_q^{pl}$ is the plasmon-dispersion, and $|a_q^{pl}|^2$ is the electron-plasmon coupling matrix element. To get $\omega_q^{pl}$ and $|a_q^{pl}|^2$, we evaluate $\varepsilon_q(\omega)$, $W_q(\omega)$, and the EELS spectrum defined Eq.~(\ref{eq:EELS}) first. Afterwards, we extract the plasmon dispersion $\omega_q^{pl}$ by identifying the maximum of $EELS_q(\omega)$ for every $q$. This allows us finally to calculate the electron-plasmon coupling $|a_q^{pl}|^2$ via
	\begin{align}
		|a_q^{pl}|^2 = \frac{\omega_q^{pl}}{2}
			   \left[ 
				   W(q, \infty) - W(q, 0)
			   \right].
	\end{align}
	The resulting dispersion reproduces all features described in the main text. From the approximate analytic solution to $\varepsilon(\omega_q^{pl}, q) = 0$ \cite{bill_electronic_2003},
	\begin{align}
		\omega_q^{pl} =
		q v_F \sqrt{ 1 + \frac{ (N_0 U_q)^2 }{0.25 + N_0 U_q} }, \label{eq:WPl}
	\end{align}
	we furthermore understand why lowering $N_0$ decreases $\omega_q^{pl}$. As shown in the Supplemental Material, the extraction of the electron-plasmon coupling reproduces the analytic expression $|a_q^{pl}|^2 = \omega_q^{pl} U_q / 2$ \cite{caruso_theory_2016} extremely well for those $q$ points where $\omega_q^{pl}$ is sufficiently separated from the particle-hole continuum. 
	
	\textbf{Electronic Self Energy in Plasmon-Pole Approximation.} The electronic self energy in its $GW$ approximation is given by
	\begin{align}
		\Sigma_k(i\omega_n) = 
			\frac{1}{\beta}
			\sum_{qm} G_q(i\omega_m) W_{k-q}(i\omega_n-i\omega_m),
	\end{align}
	where $\beta$ is the inverse temperature, $\omega_n$ are Matsubara frequencies, and $G_q$ is the electronic propagator. By using the single plasmon pole approximation from above to describe the fully screened Coulomb interaction $W_q(\omega)$ and introducing the Nambu space \cite{nambu_quasi-particles_1960}, we obtain three anomalous  $\Sigma$ contributions, stemming from the three parts in Eq.~(\ref{eq:PPM}). In detail, we obtain two static self energy contributions
	\begin{align}
		\Sigma_k^{(1)} = \sum_q n_F(\varepsilon_q) W_q(0) 
		\quad \text{and} \quad
		\Sigma_k^{(2)} = \sum_q n_F(\varepsilon_q) \frac{2 |a_q^{pl}|^2}{\omega_q^{pl}}, \notag
	\end{align}
	and a dynamic, plasmon-induced term,
	\begin{align}
		\Sigma_k^{dyn}(i\omega_n) = 
			\frac{1}{\beta}
			\sum_{qm} 
			G_q(i\omega_m) |a_q^{pl}|^2 D_q^{pl}(i\omega_n-i\omega_m),
	\end{align}
	with $D_q^{pl}(i\omega_n) = 2\omega_q^{pl} / [(i\omega_n)^2 - (\omega_q^{pl})^{2}]$ being the bosonic (plasmonic) propagator.

	\textbf{Solving the SC-DFT Gap Equation.} To find $T_c$ we use the linearized SC-DFT gap equation from Eq.~(\ref{eq:Gap}) reformulated as a an eigenvalue problem \cite{lathiotakis_density_2004},
	\begin{align}
		\underline{\underline{\tilde{K}}} \, \underline{\Delta} = \zeta \underline{\Delta},
	\end{align}
	using the generalized kernel matrix 
	\begin{align}
		\underline{\underline{\tilde{K}}}_{\xi, \xi'} =
		\begin{cases}
			-Z(\xi) - \frac{1}{2} N(\xi) K(\xi, \xi) \frac{\operatorname{tanh}\left[(\beta/2)\xi \right]}{\xi} & \xi = \xi' \\
			- \frac{1}{2} N(\xi') K(\xi, \xi') \frac{\operatorname{tanh}\left[(\beta/2)\xi' \right]}{\xi'} & \text{otherwise}
		\end{cases}. \notag
	\end{align}
	From this we find $T_c$ as the temperature at which the leading eigenvalue is $\zeta = 1$. We use an Einstein phonon with given frequency $\omega^{ph}$ and electron-phonon coupling $g^2$ yielding $\alpha^2F_{ph}(\omega) = N_0 g^2 \delta(\omega^{ph}-\omega)$. The plasmonic properties are rendered by $\alpha^2F_{pl}(\omega)$. 
	
	It is well known that the electron-phonon coupling renormalizes the electronic spectrum within the Debye window around the Fermi level. Thereby, it enhances the effective electron mass, which needs to be considered in form of the phononic $Z^{ph}(\xi)$ in the superconducting state. In the case of plasmonic excitations and their coupling to the electronic states a corresponding shift of spectral weight is also expected \cite{ayral_spectral_2012, werner_dynamical_2016}, but has so far been neglected within SC-DFT treatments. As we show in the Supplemental Material, neglecting this plasmonic contribution to $Z(\xi)$ is indeed a reasonable approximation for 3D dispersion-less plasmons with energies on the order of the electronic band width. However, for the case of 2D gapless plasmons there is a significant effect on the electronic spectrum around the Fermi energy, and we need to take $Z^{pl}(\xi)$ into account. Otherwise, we would overestimate the plasmonic contributions to the critical temperature. Doing so, the mass renormalization function $Z(\xi) = Z^{ph}(\xi) + Z^{pl}(\xi)$ is given by
	\begin{align}
		Z^{ph}(\xi) = 
		    g^2
			\int d\xi' \,
			N(\xi')
			\frac{ I_Z(\xi,\xi',\omega^{ph}) - 2J_Z(\xi,\xi',\omega^{ph}) }{\operatorname{tanh}\left[ (\beta/2) \xi \right]}
	\end{align}
	and
	\begin{align}
		Z^{pl}(\xi) = 
			&\int d\omega \, \alpha^2F_{pl}(\omega) \ \times \\ 
			&\int d\xi' \, \frac{N(\xi')}{N_0} 
			\frac{ I_Z(\xi,\xi',\omega) - 2J_Z(\xi,\xi',\omega) }{\operatorname{tanh}\left[ (\beta/2) \xi \right]}. \notag
	\end{align}
	Here, we use the $Z^{ph}(\xi)$ definition from Ref.~\cite{akashi_density_2013} also for $Z^{pl}(\xi)$ with a minor modification in the involved function $p(x) = [\operatorname{tanh}(20 \beta x)]^2$ which strongly stabilized the convergence, here (see reference for the definition of $I_Z$ and $J_Z$). 
	The full SC-DFT kernel is given by $K(\xi,\xi') = K^{ph}(\xi,\xi') + K^{stat}(\xi,\xi') + K^{dyn}(\xi,\xi')$ \cite{akashi_high-temperature_2012,luders_ab_2005}, with
	\begin{align}
		K^{ph}(\xi, \xi') = 	
			&2g^2
			\frac{
				I_K(\xi, \xi', \omega^{ph})
				-
				I_K(\xi,-\xi', \omega^{ph})
			}{
			\operatorname{tanh}\left[ (\beta/2) \xi \right]
			\operatorname{tanh}\left[ (\beta/2) \xi' \right]
			} \\
		K^{stat}(\xi, \xi') = &\frac{\mu}{N_0} \\
		K^{dyn}(\xi,\xi')   = &\frac{\lambda^{pl}}{N_0} + \Delta K^{dyn}(\xi,\xi') \\
		\Delta K^{dyn}(\xi,\xi') =
				&\frac{1}{N_0}
				\int d\omega \, 2 \AFPl \ \times \notag \\
				&     \frac{
					I_K(\xi, \xi', \omega)
					-
					I_K(\xi,-\xi', \omega)
				}{
				\operatorname{tanh}\left[ (\beta/2) \xi \right]
				\operatorname{tanh}\left[ (\beta/2) \xi' \right]
			}.
	\end{align}
	While the phononic frequencies $\omega^{ph}$ and the effective electron-phonon couplings $g^2$ are meant to be adjustable constants, the parameters connected to the Coulomb interaction, i.e. $\mu$ and $\lambda^{pl}$ are calculated from the model according to Eq.~(1) from Ref.~\cite{schonhoff_interplay_2016} and Eq.~(\ref{eq:LambdaPlasmon}) from the main text, respectively.
	
	The dynamic Coulomb Kernel $K^{dyn}(\xi,\xi')$ is derived from the Fermi surface average of $K^\text{dyn}_{k,k-q}$, as given by Akashi and Arita in Eq.~(4) of Ref.~ \cite{akashi_development_2013}, which reads in momentum space and using the single plasmon-pole approximation
	\begin{align}
		K^\text{dyn}_{k,k-q} = 
		& 2 |a_q^{pl}|^2 \ \times \\
		&	\left( 
		\frac{1}{\omega_q^{pl}}
		+
		\frac{
			I_K(\xi_k, \xi_{k-q}, \omega_q^{pl}) 
			- I_K(\xi_k,-\xi_{k-q}, \omega_q^{pl})
		}
		{
			\operatorname{tanh}\left[ (\beta/2) \xi_k  \right]
			\operatorname{tanh}\left[ (\beta/2) \xi_{k-q} \right]
		}
		\right). \notag
	\end{align}
	Here, $\xi_k$ describes the electronic dispersion and the $I_K$ function is given in Eq.~(55) in Ref.~\cite{luders_ab_2005}.
	
	To verify our implementation we present some benchmark calculation in the Supplemental Material.
	
	\textbf{Computational Parameters.} For the evaluation of the bare Coulomb interaction we use $\gamma = 1.5\,$\AA$^{-1}$ and $\varepsilon_\infty = 25$ as well as $d = 5\,$\AA\ to evaluate $\varepsilon_{rest}(q)$. The polarization function $\Pi_q(\omega)$ is calculated on a $120 \times 120$ $q$-grid based on a $120 \times 120$ $k$-grid involving $400$ frequency points between $0$ and $1\,$eV. The broadening parameter $i0^+$ is set to $10\,$meV ($20\,$meV and $50\,$meV result in identical trends, however, with slightly reduced electron-plasmon couplings). The plasmonic Eliashberg function is evaluated on a frequency grid using $800$ points between $0$ and and $1\,$eV. All $\delta(x)$ functions are approximated by Gaussian functions with a smearing of $7\,$meV. The SC-DFT gap equation is solved on an energy grid ranging from the lower to the upper end of the band width. We use $1200$ grid points which are distributed logarithmically within a window of $\pm0.1\,$eV around the Fermi level ($3/4$ of all points). All other points are distributed linearly.

	\section{Acknowledgement}	
	
	S.H. and M.R. acknowledge support from DOE under Grant No. DE-FG02-05ER46240. M.R. would like to thank the Alexander von Humboldt Foundation for support. T. W., G. S., and J.B. acknowledge support from DFG via RTG 2247 as well as the European Graphene Flagship. Numerical computations were carried out on the University of Southern California high-performance supercomputer cluster and the North-German Supercomputing Alliance (HLRN) cluster.

	\section{Supplemental Material to: Plasmonic Superconductivity in Layered Materials}
	
	\section{Plasmonic Properties}
	
\begin{figure*}
		\includegraphics[width=0.32\textwidth]{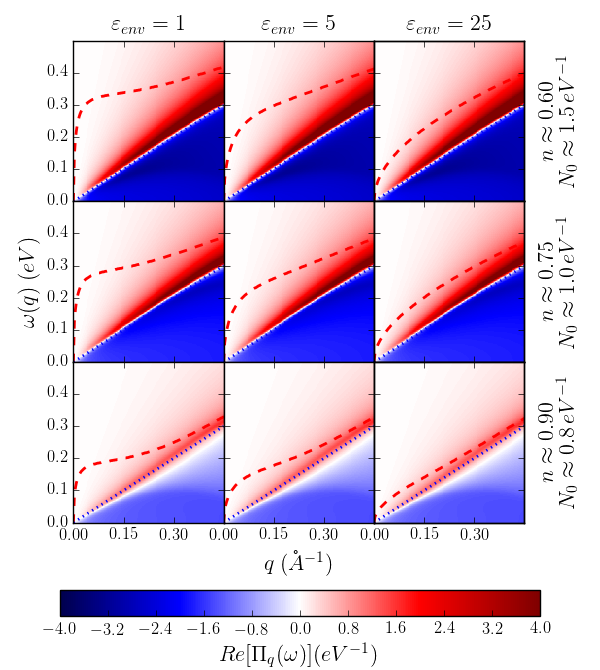}
		\includegraphics[width=0.32\textwidth]{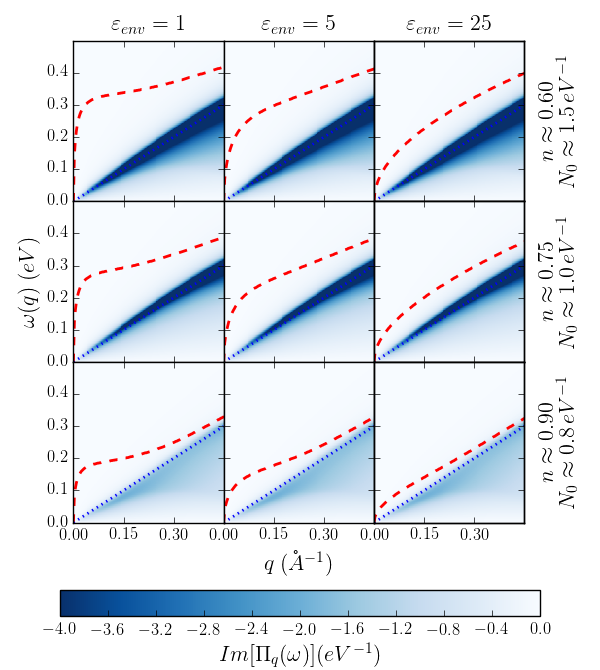}
		\includegraphics[width=0.32\textwidth]{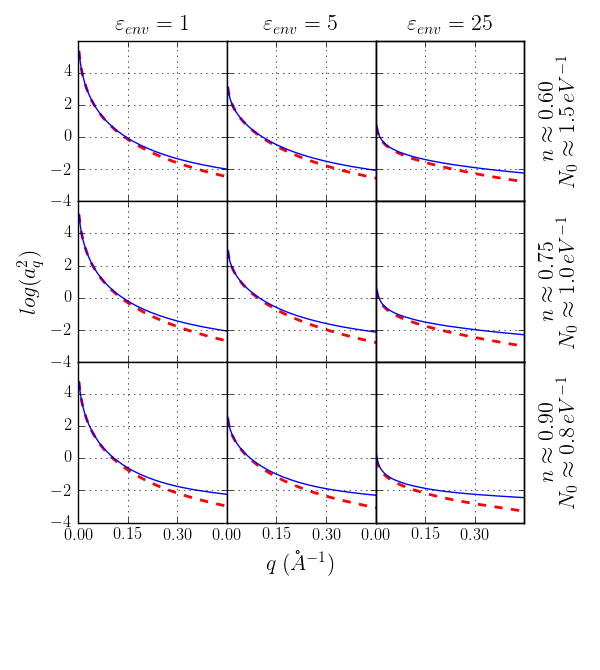}
		\caption{Left and right panels: Real and imaginary parts of the polarization functions $\Pi(q,\omega)$ (colormaps) and plasmon frequencies (dashed lineds). Right panels: Electron-plasmon coupling $|a_q^{pl}|^2$ extracted from fits (red dashed) and calculated from model (blue solid). \label{figSupp:Pol}}
	\end{figure*}

	In Fig.~\ref{figSupp:Pol} we show the real and imaginary parts of the polarization function $\Pi_q(\omega)$ and the electron-plasmon coupling $|a_q^{pl}|^2$ for different screening and doping levels. From the imaginary parts we see that there is no Landau damping in the regions where the plasmonic dispersion displays a shoulder. In these regions the numerically extracted/fitted electron-plasmon coupling coincide with the analytic expressions ($|a_q^{pl}|^2 = 0.5 w_q^{pl} U_q$) from Ref. \cite{caruso_theory_2016}. As soon as the plasmon frequencies get closer to the continuum, the fitted and analytic values for $|a_q^{pl}|^2$ start to differ and the fitted electron-plasmon coupling becomes smaller than the analytical results.
	
	\begin{figure}
		\includegraphics[width=0.99\columnwidth]{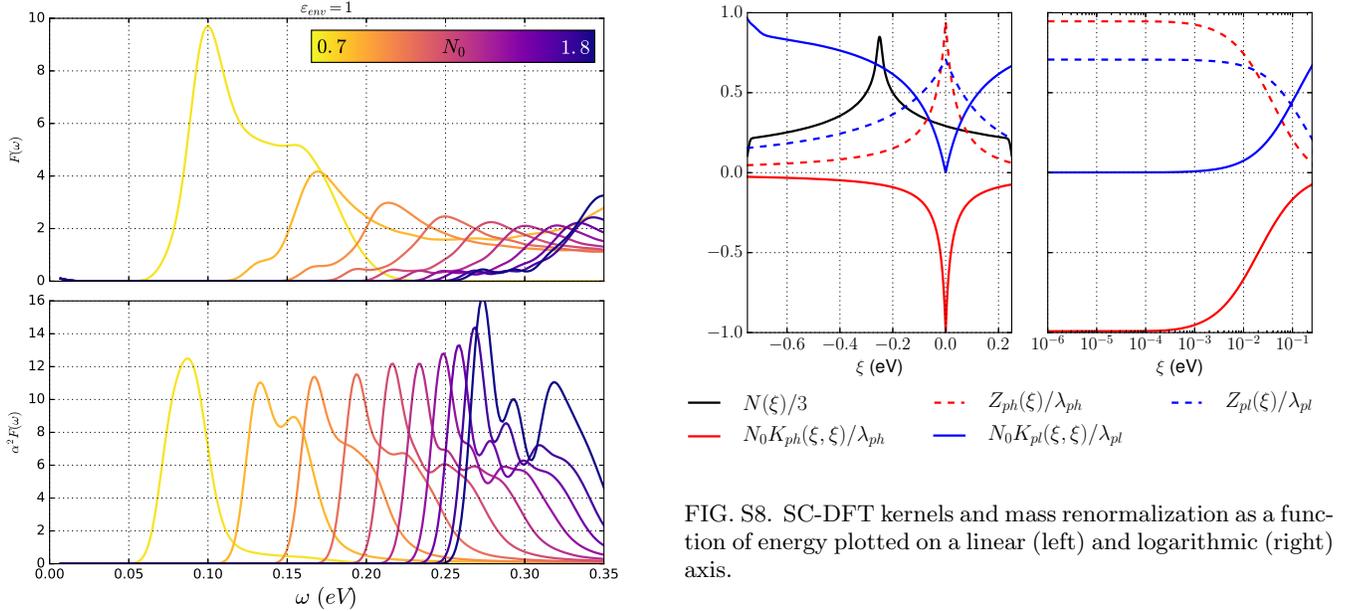}
		\caption{Plasmonic DOS (top) and full plasmonic Eliashberg function (bottom) for different doping levels. \label{figSupp:FA2F}}
	\end{figure}
	
	In Fig. \ref{figSupp:FA2F} we show $\alpha^2F_{pl}(\omega)$ and $F_{pl}(\omega)$. We noted in the main text that $\alpha^2F_{pl}(\omega)$ displays only a shift in position with a change in doping, while the peak height remains roughly constant. The shift is explained by the strong reduction of the plasmon frequencies ($\omega_q$) for decreasing $N_0$, as can be seen in the left panel of Fig. \ref{figSupp:Pol}. In the top panel of Fig. \ref{figSupp:FA2F} we observe that $F_{pl}(\omega)$ shifts to smaller energies while simultaneously increasing in strength due to a flattening of $\omega_q$ with decreasing $N_0$ (see left panel of Fig. \ref{figSupp:Pol}). This overcompensates the reduction of $\alpha^2F_{pl}(\omega)$ due to decreasing $|a_q^{pl}|^2 = 0.5 w_q^{pl} U_q$ with decreasing $w_q^{pl}$.
	
	\section{SC-DFT Benchmarks}
	
	In order to benchmark our single-band SC-DFT implementation, we performed several tests, comparing results from our code to available data and known literature. 
	
	\subsection{Kernels}
	
	\begin{figure}
		\includegraphics[width=0.99\columnwidth]{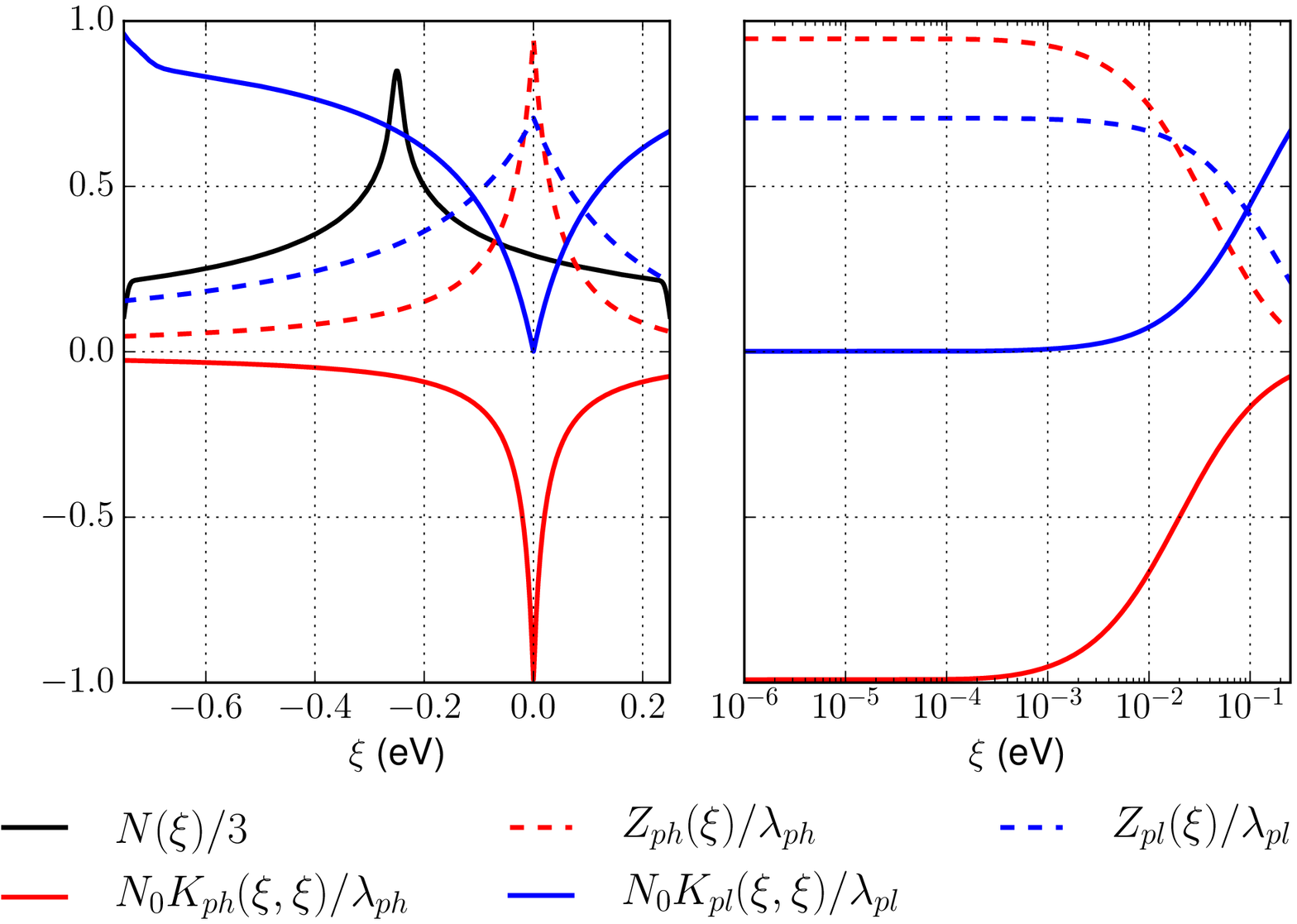}
		\caption{SC-DFT kernels and mass renormalization as a function of energy plotted on a linear (left) and logarithmic (right) axis. \label{fig:Benchmarkernels}}
	\end{figure}
	
	First, we ensure that the numerically evaluated kernels ($K^{ph}$ and $K^{pl}$) and mass renormalization functions ($Z^{ph}$ and $Z^{pl}$) satisfy the approx. limits (under the assumption that the characteristic boson frequencies are small compared to the electronic band widths) described in the main text:
	$$ N_0 K^{ph}(0,0) \approx -\lambda^{ph}, \hspace{1cm} Z^{ph}(0) \approx \lambda^{ph},$$
	$$ N_0 K^{pl}(0,0) \approx 0, \hspace{1cm} Z^{pl}(0) \approx \lambda^{pl}.$$
	
	In Fig. \ref{fig:Benchmarkernels} we show all four functions. Both kernels behave as expected, while the $Z$ functions show small deviations from the above limits due to two reasons. First, we have an asymmetric density of states around the Fermi level, which forces us to use the asymmetrical definition of $Z$ discussed in Ref. \cite{akashi_density_2013} and which lowers $Z(\xi)$. Second, considering the $\xi \rightarrow 0$ limit of $Z(\xi)$,
	\begin{align}
	Z(0) = 
	&\int d\omega \, \alpha^2F(\omega) \ \times \\ 
	\lim_{\xi \rightarrow 0}&\int d\xi' \, \frac{N(\xi')}{N_0} 
	\frac{ I_Z(\xi,\xi',\omega) - 2J_Z(\xi,\xi',\omega) }{\operatorname{tanh}\left[ (\beta/2) \xi \right]}. \notag
	\end{align}
	we need to fulfill
	\begin{align}
	\lim_{\xi \rightarrow 0}
	\int d\xi' \frac{N(\xi')}{N(0)}\frac{I_Z(\xi,\xi',\omega) - 2J_Z(\xi,\xi',\omega)}{\tanh[(\beta/2)\xi]} = \frac{2}{\omega}.
	\end{align}
	in order to satisfy $Z(0) = \lambda = \int dw \frac{2\alpha^2F(\omega)}{\omega}$.
	
	The decays of the $I_Z$ and $J_Z$ functions are determined by the characteristic frequency of the bosonic modes involved (phonons and plasmons). In the phonon case, the decay energy interval is small compared to the electronic band width ($Z^{ph}$ goes to zero rapidly), and the above requirement is easily met, hence $Z^{ph}(0) \approx \lambda^{ph}$. In the plasmonic case, however, the characteristic frequency is much higher (on the order of 100 meV). Hence the decay energy interval is much larger, which results in a larger discrepancy between the observed and expected limits. In this case, the above integral is cut off when $N(\xi') = 0$, and not by the decay of the $I_Z$ and $J_Z$ functions. This effect has previously been discussed by Arita and Akashi \cite{akashi_fullerides_2013} in the use of SC-DFT for materials with narrow bandwidths, and results in a slightly increased transition temperature.
	
	\subsection{Step-Like Density of States}
	
	\begin{figure}
		\includegraphics[width=0.99\columnwidth]{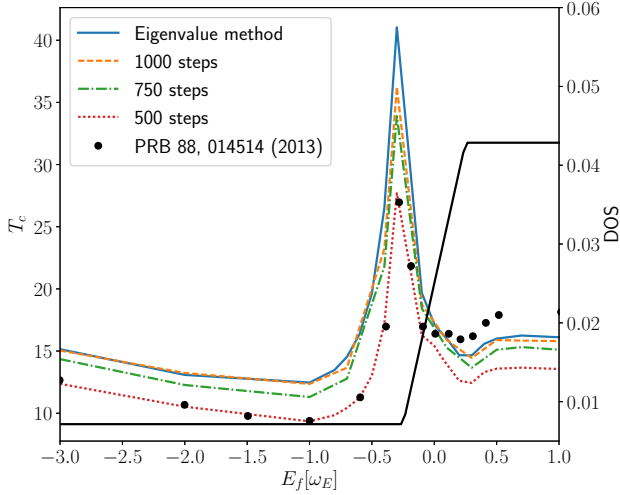}
		\caption{SC-DFT Benchmark I. Comparison between our $T_c$ (lines) and data published in Ref.~\cite{akashi_density_2013} (black dots) for a step-like density of states (gray line) including the effects of electron-phonon coupling and static electron-electron repulsion. We show results obtained by using the iterative method to solve the gap equation with different numbers of steps allowed before terminating the loop (dashed lines) and by using the eigenvalue method (solid line). The eigenvalue method appears as the limit of an increasing number of iteration steps. \label{fig:BenchmarkArita}}
	\end{figure}
	
	Next we compare our results to those obtained by Akashi and Arita published in Ref.~\cite{akashi_density_2013}. In analogy to this reference, we use an Einstein phonon with frequency $\omega_E \approx 50\,$meV with an effective electron-phonon coupling of $\lambda = 1$ and an effective electron-electron repulsion of $\mu = 0.5$. The underlying density of states is given by a step-like function, as described in Ref.~\cite{akashi_density_2013} with a band-width of $40\,$eV. Akashi and Arita determined $T_c$ by self-consistently solving the gap equation and setting $T_c$ to the highest temperature for which they could find a non-trivial solution. Such an implementation requires the use of a parameter that limits the number of allowed iterations to obtain a solution to the gap equation before terminating the process. In our implementation we cast the gap equation into an eigenvalue problem and find $T_c$ as the temperature for which the leading eigenvalue is one. In Fig.~\ref{fig:BenchmarkArita} we show results we obtained using both methods (showing the difference due to changes in the parameter) along with the results from Ref.~\cite{akashi_density_2013}. Besides decreased critical temperatures $T_c$ above $E_f = 0$ the agreement between our and the reference data is very good for the case where $500$ iterations steps are allowed. Like in Ref.~\cite{akashi_density_2013} we find a maximum of $T_c$ in the vicinity of $E_F \approx -0.3 \omega_E$ and similar trends above and below this point. The minor differences between the two data sets trace back to slightly different energy grids, bandwidth cut-offs, and different tolerance settings.

	\subsection{MoS$_2$ Monolayer}
	
	As a second benchmark, we compare critical temperatures obtained by evaluating the Allen-Dynes equation, using input data from density functional perturbation theory and from our SC-DFT implementation for electron-doped monolayers of MoS$_2$. We use the effective frequencies $\omega_{log} = \omega^{ph}$ and effective electron-phonon couplings $\lambda^{ph}$ from Ref.~\cite{rosner_phase_2014} to construct Einstein-phonon models for a range of different doping levels. Together with the electronic dispersion of the occupied band of MoS$_2$, which we obtain from a corresponding Wannier construction, we can solve the SC-DFT gap equation given in Eq.~(7) of the main text. To this end, we use similar $k$ meshes and broadenings as in Ref.~\cite{rosner_phase_2014} and fix the resulting $\mu^*=0.15$. In Tab.~\ref{tab:BenchmarkMoS2} we list the resulting critical temperatures from the reference ($T_c^\text{Allen-Dynes}$) and the solution of the SC-DFT gap equation ($T_c^\text{SC-DFT}$). For $\lambda^{ph} < 2$ we find similar critical temperatures from both approaches. For increased effective electron-phonon couplings, the resulting $T_c$ differ, as the Allen-Dynes equation is known to underestimate $T_c$ in the strong-coupling regime \cite{allen_transition_1975}. Apart from this, small differences occur due to slightly different bandwidths and the applied rigid-shift approximation to describe the doping in the $T_c^\text{SC-DFT}$ data. Considering these circumstances, the agreement between both approaches is good. 
	
	\begin{table}[ht]
		\begin{threeparttable}
			\caption{ SC-DFT Benchmark II. Comparison of critical temperatures for electron-doped monolayers of MoS$_2$, obtained by evaluating the Allen-Dynes equation using density functional perturbation theory input from Ref.~\cite{rosner_phase_2014} and by solving the energy-dependent SC-DFT gap equation. \label{tab:BenchmarkMoS2} }
			\begin{ruledtabular}
				\begin{tabular}{llllll}
					doping\tnote{a}         
					& $0.075$ & $0.087$ & $0.100$ & $0.112$ & $0.125$ \\ \hline
					$\lambda^{ph}$ 
					& $0.821$ & $1.236$ & $1.920$ & $3.089$ & $7.876$ \\
					$\omega^{ph}$ [eV]
					& $0.029$ & $0.025$ & $0.021$ & $0.016$ & $0.009$ \\
					$N_0$ [eV$^{-1}$]\tnote{b}         
					& $0.440$ & $0.569$ & $0.690$ & $0.857$ & $0.895$ \\ \hline
					$T_c^\text{Allen-Dynes}$ [K]    
					& $8.01$  & $15.57$ & $21.12$ & $22.11$ & $15.51$ \\
					$T_c^\text{SC-DFT}$ [K]
					& $5.40$  & $12.76$ & $22.44$ & $35.07$ & $39.60$ 
				\end{tabular}
			\end{ruledtabular}
			\begin{tablenotes}
				\item[a]{\footnotesize{given in additional electrons per unit cell}}
				\item[b]{\footnotesize{given in states/spin/eV/unit cell}}
			\end{tablenotes}
		\end{threeparttable}
	\end{table}
	
	\section{Interacting Spectral Function}
	
	\begin{figure*}
		\includegraphics[width=0.42\textwidth]{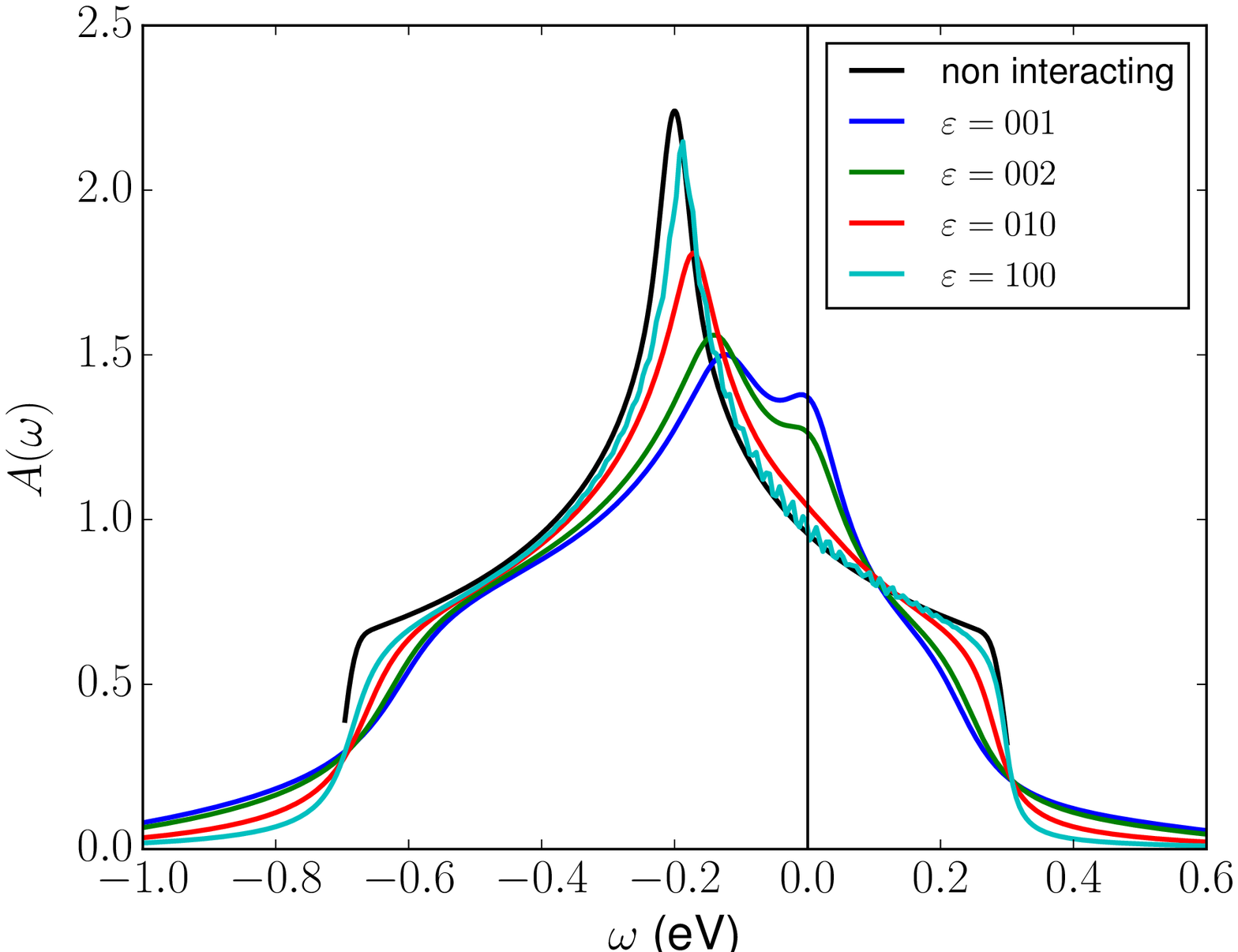}
		\includegraphics[width=0.42\textwidth]{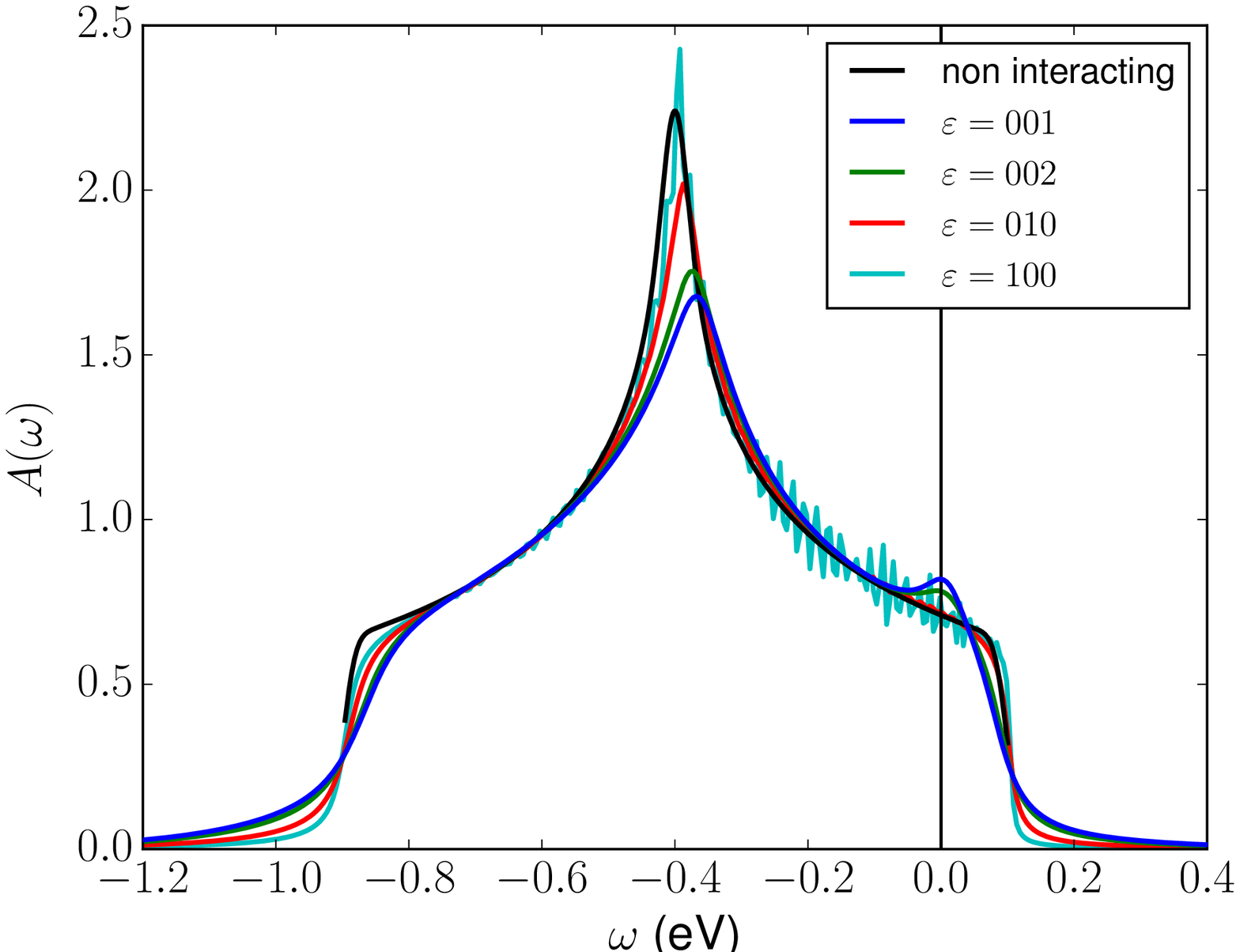}
		\caption{Interacting spectral functions from electron-plasmon self energies. The substrate screening dependencies to the electron-plasmon renormalization of the spectral function (colored lines) is shown along with the non-interacting density of states (black) for different doping levels as indicated by the Fermi level (vertical line).\label{figSupp:Spectral}}
	\end{figure*}
	
	We check the influence of the electron-plasmon interaction on the spectral function by considering the contribution of the dynamical Coulomb interaction on the electron self energy only, using Eq.~(3) from the main text 
	\begin{align}
	\Sigma_k^{dyn}(i\omega_n) = 
	\frac{1}{\beta}
	\sum_{k'm} 
	G_{k}(i\omega_m) |a_q^{pl}|^2 
	\frac{2\omega_q^{pl}}{(i\omega_n-i\omega_m)^2 - (\omega_q^{pl})^{2}}.
	\end{align}
	This formulation is based on the plasmon pole approximation and allows us within the $G_0W_0$ approximation to evaluate the Matsubara sum analytically \cite{Mahan}, yielding 
	\begin{align}
	&\Sigma^{dyn}_{k}(\omega) =
	\int_{BZ} dq \,
	|a_q^{pl}|^2 \\
	&\left[
	\frac{n_q + f_{k+q}}{\omega - \xi_{k+q} + \omega_q^{pl} + i\delta^+}
	+
	\frac{n_q + 1 - f_{k+q}}{\omega - \xi_{k+q} - \omega_q^{pl} + i\delta^+}
	\right]. \notag
	\end{align}
	This corresponds to the self-energy given in Ref. \cite{caruso_theory_2016}.
	From the electron-plasmon self-energy we calculate the interacting spectral functions,
	$$ A(\omega) = \frac{1}{\pi}\sum_k \frac{|\text{Im}\Sigma^{dyn}_k(\omega)|}{[\omega - \epsilon_k - \text{Re}\Sigma^{dyn}_k(\omega)]^2 + [\text{Im}\Sigma^{dyn}_k(\omega)]^2}, $$
	as shown for different scenarios in Fig.~\ref{figSupp:Spectral}. These calculations were performed on $80 \times 80$ k-meshes and equivalent q-meshes using Gaussian functions instead of Dirac delta functions (for the energies) with $15\text{meV}$ broadening and with a value of $0.3\text{eV}$ for $\delta^+$.
	
	We observe a reduction of the band-width in combination with arising plasmonic satellites above and below the non-interacting band edges. At the same time the former van-Hove singularity shifts to higher energy and, most important, the density of states at the Fermi level is enhanced. The latter is equivalent to an effective electron mass enhancement, which is why we decided to take the plasmonic mass enhancement factor $Z^{pl}(\omega)$ into account. All of these effects are controlled by the dielectric environment and the doping level.
	
	If we compare this to the spectral fingerprints of plasmons in a 3D (semiconducting) silicon system (see Ref. \cite{caruso_spectral_2015}) we find a different scenario. In 3D, due to the non-dispersive plasmonic modes at energies around $5$ to $10\,$eV, there is just a replica of the original non-interacting band structure shifted to lower energies (in the amount of the non-dispersive plasmonic energy). This might also been seen as a plasmonic satellite. But, this satellite band structure is strongly reduced in its spectral weight since in 3D the coupling scales as $|a_q^{pl}|^2 \propto 1/q^2$, while in 2D we have $|a_q^{pl}|^2 \propto 1/q$. In 2D we thus observe two major effects: (a) low-energy plasmonic modes which result in low-energy changes to the band structure around the Fermi level and (b) an enhanced coupling of the low energy modes, which enhance the low-energy spectral finger prints. 
	
	\section{Origin of Plasmonic $T_c$ Reduction}
	
	As outlined in the main text the following holds for a plasmonic reduction of the full critical temperature:
	\begin{align}
	R = \frac{\omega^{tot}}{\omega^{ph}}
	\operatorname{exp}
	\left[ 
	\frac{1+\lambda^{ph}}{\lambda^{ph} - \mu^{ph*}}
	-\frac{1+\lambda^{tot}}{\lambda^{tot} - \mu^{tot*}}
	\right] < 1.
	\end{align}
	Within the chosen parameters the ratio $\frac{\omega^{tot}}{\omega^{ph}} > 1$ can not be responsible for $R$ being smaller than $1$. Therefore, the argument of the exponential function must me smaller than $0$,
	\begin{align}
	\frac{1+\lambda^{ph}}{\lambda^{ph} - \mu^{ph*}}
	-\frac{1+\lambda^{tot}}{\lambda^{tot} - \mu^{tot*}}
	< 0.
	\end{align}
	In the limit $\lambda^{ph} \gg \lambda^{pl}$ we can approximate $\lambda^{tot} \approx \lambda^{ph}$ and can reformulate the statement to
	\begin{align}
	\frac{1+\lambda^{ph}}{\lambda^{ph} - \mu^{ph*}}
	< \frac{1+\lambda^{ph}}{\lambda^{ph} - \mu^{tot*}},
	\end{align}
	which is true for
	\begin{align}
	\mu^{tot*} > \mu^{ph*}.
	\end{align}
	This, in turn is equivalent to
	\begin{align}
	\frac{\lambda^{ph} + \lambda^{pl}}{1 + \operatorname{ln}(E_F/\omega^{tot})} &>
	\frac{\lambda^{ph}}{1 + \operatorname{ln}(E_F/\omega^{ph})}\\
	\Leftrightarrow
	\underbrace{\frac{\lambda^{ph} + \lambda^{pl}}{\lambda^{ph}}}_{\gtrsim 1} &>
	\underbrace{\frac{1 + \operatorname{ln}(E_F/\omega^{tot})}{1 + \operatorname{ln}(E_F/\omega^{ph})}}_{< 1 \text{ since } \omega^{tot} > \omega^{ph}}.
	\end{align}
	$\mu^{tot*}$ is thus slightly enhanced in comparison to $\mu^{ph*}$ due to $\lambda^{pl}$ and the increased effective frequency $\omega^{tot}$.
	
	\bibliography{bibliography,bibliography_supplement}	

\end{document}